\documentclass[pra,twocolumn,tightenlines,showpacs,floatfix,nofootinbib]{revtex4}
\usepackage{multirow}
\usepackage{amsmath}
\usepackage{bm}
\usepackage{dcolumn}
\usepackage{graphicx}

\newcommand{\eref}[1]{Eq.~(\ref{#1})}
\newcommand{\tref}[1]{Table~\ref{#1}}

\begin{document}
\title{Long range interaction coefficients for ytterbium dimers}

\author{S.~G.~Porsev$^{1,2}$}
\author{M.~S.~Safronova$^{1,3}$}
\author{A. Derevianko$^4$}
\author{Charles W. Clark$^3$}
\affiliation{ $^1$Department of Physics and Astronomy, University of Delaware,
    Newark, Delaware 19716, USA\\
$^2$Petersburg Nuclear Physics Institute, Gatchina, Leningrad District, 188300, Russia \\
$^3$Joint Quantum Institute, National Institute of Standards and Technology and the \\
University of Maryland, Gaithersburg, Maryland, 20899, USA
$^4$Physics Department, University of Nevada, Reno, Nevada 89557, USA}
\date{\today}

\begin{abstract}
We evaluate the electric-dipole and electric-quadrupole static
and dynamic polarizabilities for the $6s^2\,^1\!S_0$, $6s6p\,^3\!P^o_0$, and
$6s6p\,^3\!P^o_1$ states and estimate their uncertainties.
A methodology is developed for an accurate evaluation of the van der Waals coefficients of dimers
involving excited state atoms with strong decay channel to the ground state.
This method is used for evaluation of the long range interaction
coefficients of particular experimental interest, including the
 $C_6$ coefficients for the Yb--Yb $^1\!S_0+\,^3\!P^o_{0,1}$ and $^3\!P^o_0+\,^3\!P^o_{0}$ dimers and
$C_8$ coefficients for the $^1\!S_0+\,^1\!S_0$ and $^1\!S_0+\,^3\!P^o_1$ dimers.
\end{abstract}
\maketitle

\section{Introduction}
The ytterbium atom has two fermionic and five bosonic isotopes, a $^1\!S_0$ ground state, a long-lived metastable $6s6p\,\, ^3\!P_0^o$ state, and transitions at convenient wavelengths for laser cooling and trapping. All this makes Yb a superb candidate for a variety of applications such as
development of optical atomic clocks~\cite{Ybclockl}, study of degenerate quantum gases~\cite{TakKomHon04l}, quantum information
processing~\cite{GorReyDal09l}, and studies of fundamental symmetries~\cite{PNC1l}.
The best limit to date on the value of the electron electric-dipole moment (EDM) which constrains extensions of the standard model of
electroweak interactions, was obtained using the YbF molecule~\cite{HudKarSma11l}.
YbRb and YbCs molecules have also been proposed for searches for the electron EDM~\cite{MeyBoh09} since they
can be cooled to very low temperatures and trapped in optical dipole traps, leading to very long coherence times in
comparison to molecular beam EDM experiments.

Yb is of particular interest for studying quantum gas
mixtures \cite{BorCiuJul09l,TakKomHon04l,TojKitEno06l,EnoKitToj08,KitEnoKas08l,YamTaiSug13l,TakSaiTak12l,NemBauMun09l,MunBruMad11,BauMunGor11l}.
Significant progress has been achieved in studying the properties of Yb-Yb photoassociation spectra at ultralow
temperatures~\cite{BorCiuJul09l}. Photoassociation spectroscopy has been performed on
bosons~\cite{TakKomHon04l,TojKitEno06l} and fermions~\cite{EnoKitToj08}.
The use of optical Feshbach resonances for control of entangling interactions between nuclear
spins of $^{171}$Yb atoms for quantum information processing applications has been proposed in~\cite{ReiJulDeu09}.
A p-wave optical Feshbach resonance  using purely long-range molecular states of a
fermionic isotope of ytterbium $^{171}$Yb was demonstrated in~\cite{YamTaiSug13l}.
Recent work~\cite{SanOdoJav11} theorizes that the case of $^{174}$Yb may have sufficiently small direct background interaction
between the atoms to support two bound states that represent attractively and repulsively bound dimers occurring simultaneously.

The excited molecular states asymptotically connected to the $^1\!S_0 +\, ^3\!P^o_1$
separated Yb atom limit were investigated by Takasu {\it et. al.} in~\cite{TakSaiTak12l}.
They reported the successful production of a subradiant $1_g$ state of a
two-atom Yb system in a three-dimensional optical lattice. The properties of the long-range
potential were studied and the van der Waals coefficients $C_3$, $C_6$, and $C_8$ were predicted.
However,  fit of the $C_6$ and $C_8$ coefficients for the $1_g$ state was rather
uncertain, with strong correlation between the $C_6$ and $C_8$ fit parameters~\cite{Jul13}.

Knowledge of the $C_6$ and $C_8$ long-range interaction coefficients in Yb-Yb
dimers is critical to understanding the physics of dilute gas mixtures.
Recently, we evaluated the $C_6$ coefficient for the Yb-Yb $^1\!S_0 +\,^1\!S_0$ dimer and
found it to be $C_6=1929(39)$~\cite{SafPorCla12}, in  excellent agreement with the experimental
result $C_6=1932(35)$~\cite{KitEnoKas08l}. However, the same method cannot be directly applied to
the calculation of the van der Waals coefficients with Yb-Yb $^1\!S_0+\,^3\!P_1^o$ dimer owing to
the presence of the $^3\!P_1^o\rightarrow\,^1\!S_0$ decay channel.

In this work, we develop the methodology for an accurate evaluation of the van der Waals coefficients of dimers
involving excited state atoms with a strong decay channel to the ground state and evaluate $C_6$ and $C_8$
coefficients of  particular experimental interest.
We carefully study the uncertainties of all quantities calculated in this work
so the present values can be reliably used to analyse existing measurements and to facilitate planning of the future
experimental studies.
The methodology developed in this work can be used for evaluation of van der Waals coefficients in a variety of systems.
\section{General formalism}
\label{GenForm}
We investigate the molecular potentials asymptotically connecting to
the $|A\rangle + |B\rangle$ atomic states. The wave function of such a system
constructed from these states is
\begin{equation}
|M_A,M_B;\Omega \rangle =|A\rangle_{\mathrm{I}}\, |B\rangle_{\mathrm{II}},
\label{Yb_WF}
\end{equation}
where the index I(II) describes the wave function located on the center I(II) and
$\Omega =M_A+M_B$. Here, the $M_{A(B)}$ is the projection of the appropriate total atomic angular
momentum ${\bf J}_{A(B)}$ on the internuclear axis. We assume that $\Omega$ is
a good quantum number for all calculations in this work (Hund's case (c)).

The molecular wave functions can be obtained by diagonalizing the molecular
Hamiltonian
\begin{equation}
\hat{H}=\hat{H}_A + \hat{H}_B+\hat{V}(R)
\label{Eq_Hamilt}
\end{equation}%
in the model space. Here, $\hat{H}_A$ and $\hat{H}_B$ represent the Hamiltonians of the two noninteracting
atoms and $\hat{V}(R)$ is the residual electrostatic potential defined as the
full Coulomb interaction energy in the dimer excluding interactions of the
atomic electrons with their parent nuclei.

Unless stated otherwise, throughout this paper we use atomic units (a.u.);
the numerical values of the elementary
 charge, $|e|$, the reduced Planck constant, $\hbar = h/2
\pi$, and the electron mass, $m_e$, are set equal to 1.
The atomic unit for polarizability can be converted to SI units via
$\alpha/h$~[Hz/(V/m)$^2$]=2.48832$\times10^{-8}\alpha$~(a.u.), where the conversion coefficient is $4\pi \epsilon_0
a^3_0/h$, $a_0$ is the Bohr radius and $\epsilon_0$ is the dielectric constant.

The potential $V(R)$ may be expressed as an expansion in the multipole
interactions:
\[
V(R)=\sum_{l,L=0}^{\infty }V_{lL}/R^{l+L+1}\,,
\]%
where $V_{lL}$ are given by~\cite{DalDav66}
\begin{eqnarray}
V_{lL}(R)&=&\sum_{\mu =-l_s}^{l_s}\frac{(-1)^{L}(l+L)!}{\left\{ (l-\mu
)!\,(l+\mu )!\,(L-\mu )!\,(L+\mu )!\right\} ^{1/2}} \nonumber \\
&\times& \left( T_{\mu}^{(l)}\right) _{I}\left( T_{-\mu }^{(L)}\right)_{II}.
\label{VlL}
\end{eqnarray}%
Here, $l_s=\textrm{min}(l,L)$ and the multipole spherical tensors
 are
\begin{equation}
T_{\mu }^{(K)}=-\sum_{i}r_{i}^{K}C_{\mu }^{(K)}(\hat{\mathbf{r}}_{i})\,,
\end{equation}%
where the summation is over atomic electrons, ${\mathbf{r}}_{i}$ is the
position vector of electron $i$, and $C_{\mu }^{(L)}(\hat{\mathbf{r}}_{i})$
are the reduced spherical harmonics~\cite{VarMosKhe88}.

We now restrict our consideration to the dipole-dipole and
dipole-quadrupole interactions.
Introducing designations
$d_{\mu} \equiv T_{\mu}^{(1)}$, $Q_{\mu} \equiv T_{\mu}^{(2)}$,
$V_{dd} \equiv V_{11}/R^3$, and $V_{dq} \equiv V_{12}/R^4$,
we obtain from~Eq.~(\ref{VlL}):
\begin{eqnarray}
V_{dd}(R) &=&-\frac{1}{R^{3}}\sum_{\mu =-1}^{1}w_{\mu }^{(1)}(d_{\mu})_{I}(d_{-\mu })_{II}, \\
V_{dq}(R) &=& \frac{1}{R^{4}}
\sum_{\mu =-1}^{1}w_{\mu }^{(2)}
\left[ (d_{\mu})_{I}(Q_{-\mu })_{II}-(Q_{\mu })_{I}(d_{-\mu })_{II}\right] , \nonumber
\label{Vdd}
\end{eqnarray}%
where the dipole and quadrupole weights are
\begin{eqnarray}
w_{\mu }^{(1)} &\equiv& 1+\delta _{\mu 0} , \nonumber \\
w_{\mu }^{(2)} &\equiv& \frac{6}{\sqrt{\left( 1-\mu \right) !\left( 1+\mu
\right) !\left( 2-\mu \right) !\left( 2+\mu \right) !}}.
\label{wmu}
\end{eqnarray}%
Numerically, $w_{-1}^{(2)}=$ $w_{+1}^{(2)}=\sqrt{3}$ and $w_{0}^{(2)}=3$.

The energy $\mathcal{E} \equiv E_A+E_B$, where $E_A$ and $E_B$ are
the atomic energies of the $|A\rangle$ and $|B\rangle$ states, is obtained from
\begin{eqnarray}
\left(\hat{H}_A + \hat{H}_B \right) |M_A,M_B;\Omega \rangle =
\mathcal{E} \, |M_A,M_B;\Omega \rangle .
\end{eqnarray}
The molecular wave function $\Psi^{g\!/u}_\Omega$ can be formed as
a linear combination of the wave functions given by~Eq.~(\ref{Yb_WF}).
$\Psi^{g\!/u}_\Omega$ poses a definite gerade/ungerade symmetry and definite
quantum number $\Omega$. It can be represented by
\begin{eqnarray}
\Psi^p_\Omega = \left\{ \begin{array}{c}
   \frac{1}{\sqrt{2}} ( |A \rangle_{\mathrm{I}}
\,|B \rangle_{\mathrm{II}} + (-1)^p |B \rangle_{\mathrm{I}} |A \rangle_{\mathrm{II}} ), \,\, A \neq B   \\
  \qquad \qquad  \quad \,\,\,
  |A \rangle_{\mathrm{I}}\,|A \rangle_{\mathrm{II}}, \qquad \qquad \quad A=B ,
\end{array}
\right.
\end{eqnarray}
where we set $p=0$ for ungerade symmetry and $p=1$ for gerade symmetry.
We have taken into account that the states $A$ and $B$ that are of interest to the present work are the
opposite parity states of Yb atom (when $A \neq B$).

Applying the formalism of Rayleigh-Schr\"{o}edinger perturbation theory in
the second order~\cite{Dal61} and keeping the terms up to $1/R^{8}$ in the expansion of $V(R)$
we obtain the dispersion potential in
two-atom basis:
\begin{eqnarray}
U(R) &\equiv &\langle \Psi _{\Omega }^{p}|V(R)|\Psi _{\Omega }^{p}\rangle \nonumber \\
&\approx &\langle \Psi _{\Omega }^{p}|\hat{V}_{dd}|\Psi _{\Omega}^{p}\rangle +
\sum_{\Psi_{i} \neq  \Psi_\Omega^p} \left[ \frac{\langle \Psi _{\Omega }^{p}|\hat{V%
}_{dd}|\Psi _{i}\rangle \langle \Psi _{i}|\hat{V}_{dd}|
\Psi _{\Omega}^{p}\rangle }{\mathcal{E}-E_{i}} \right. \nonumber \\
&+& \left. \frac{\langle \Psi _{\Omega }^{p}|\hat{V}%
_{dq}|\Psi _{i}\rangle \langle \Psi _{i}|\hat{V}_{dq}|\Psi _{\Omega
}^{p}\rangle }{\mathcal{E}-E_{i}}\right] ,
\label{UR0}
\end{eqnarray}%
The intermediate molecular state $|\Psi_{i}\rangle $ with unperturbed
energy $E_{i}$ runs over a \emph{complete} set of two-atom states, excluding
the model-space states, Eq.~(\ref{Yb_WF}).

The dispersion potential can be approximated as
\begin{equation}
U(R) \approx -\frac{C_3}{R^3}-\frac{C_6}{R^6}-\frac{C_8}{R^8} .
\label{UR1}
\end{equation}%
\subsection{First-order corrections}
The first-order correction, which is determined by the first term on the
right-hand side of~\eref{UR0}, is associated with the $C_3$ coefficient
in~\eref{UR1}. For the states considered in this work, this coefficient is nonzero only for the molecular potential asymptotically
connecting to the $^1\!S_0 +\, ^3\!P^o_1$ atomic states.
It depends entirely on the reduced matrix element (ME)
of the electric-dipole operator $|\langle ^3\!P_1^o||d||^{1}\!S_{0}\rangle|$ and is given by
a simple formula
\begin{equation}
C_{3}(\Omega _{p})=(-1)^{p+\Omega }(1+\delta _{\Omega ,0})
\frac{|\langle ^{3}\!P_{1}^{o}||d||^{1}\!S_{0}\rangle|^{2}}{3}.
\label{C3}
\end{equation}
Specifically,
\begin{eqnarray}
C_{3}(0_{g/u}) &=&\mp \,2\frac{|\langle ^{3}\!P_{1}^{o}||d||^{1}\!S_{0}\rangle |^{2}}{3}, \nonumber \\
C_{3}(1_{g/u}) &=& \pm\, \frac{|\langle ^{3}\!P_{1}^{o}||d||^{1}\!S_{0}\rangle |^{2}}{3},
\label{C3detail}
\end{eqnarray}
where the upper/lower sign corresponds to gerade/ungerade symmetry.
\subsection{Second-order corrections}
The second-order corrections, associated with the $C_6$ and $C_8$ coefficients,
are given by the second and third terms on the r.h.s. of~\eref{UR0},
\begin{eqnarray*}
-\frac{C_{6}(\Omega_p)}{R^6} &=&
\sum_{\Psi_i \neq \Psi_{\Omega_p}}
\frac{ \langle \Psi_{\Omega_p} |\hat{V}_{dd}| \Psi_{i} \rangle
       \langle \Psi_i |\hat{V}_{dd}| \Psi_{\Omega_p} \rangle} {\mathcal{E}-E_i} \nonumber \\
-\frac{C_8(\Omega _p)}{R^8} &=& \sum_{\Psi_{i} \neq \Psi_{\Omega_p}}
\frac{\langle \Psi_{\Omega_p}|\hat{V}_{dq}| \Psi_i \rangle
      \langle \Psi_i |\hat{V}_{dq}| \Psi_{\Omega_p} \rangle}{\mathcal{E}-E_i},
\end{eqnarray*}
where $\mathcal{E} = E_A + E_B$ and the complete set of doubled atomic states satisfies
the condition
\[
\sum_{\Psi_i} |\Psi _{i}\rangle \langle \Psi _{i}| = 1.
\]
After angular reduction, the $C_6$ coefficient can be expressed as
\begin{equation}
C_6(\Omega)= \sum_{J_\alpha=|J_A-1|}^{J_A+1}\, \sum_{J_\beta=|J_B-1|}^{J_B+1}
A_{J_\alpha J_\beta}(\Omega)\, X_{J_\alpha J_\beta},
\end{equation}
where
\begin{eqnarray}
&&A_{J_\alpha J_\beta}(\Omega)=  \nonumber \\
&& \sum_{\mu M_\alpha M_\beta}
\left[ w_\mu^{(1)} \left(
\begin{array}{ccc}
 J_A &  1  & J_\alpha \\
-M_A & \mu & M_\alpha
\end{array} \right)
\left(
\begin{array}{ccc}
 J_B &   1  & J_\beta \\
-M_B & -\mu & M_\beta
\end{array} \right) \right]^2,\nonumber \\
&&X_{J_\alpha J_\beta} = \sum_{\alpha,\beta \neq A,B} \frac{|\langle A||d||\alpha \rangle|^2
\, |\langle B ||d|| \beta \rangle|^2} {E_\alpha - E_A + E_\beta- E_B}
\end{eqnarray}
 with  fixed $J_\alpha$ and $J_\beta$ .

If $A$ and $B$ are the spherically symmetric atomic states and there are no downward transitions
from either of them, the $C_6$ and $C_8$ coefficients for the $A+B$ dimers are given by
well known formulas (see, e.g.,~\cite{PatTan97})
\begin{eqnarray}
C^{AB}_6 &=& C^{AB}(1,1), \nonumber \\
C^{AB}_8 &=& C^{AB}(1,2)+C^{AB}(2,1) ,
\label{vdW}
\end{eqnarray}
where the coefficients $C^{AB}(l,L)$ ($l,L=1,2$) are
quadratures of electric-dipole, $\alpha_1(i \omega)$, and electric-quadrupole,
$\alpha_2(i \omega)$, dynamic polarizabilities at an imaginary frequency:
\begin{eqnarray}
C^{AB}(1,1) &=& \frac{3}{\pi}\, \int_0^\infty\, \alpha_1^A(i \omega)\,
\alpha_1^B(i \omega)\, d\omega, \nonumber \\
C^{AB}(1,2) &=& \frac{15}{2\pi}\, \int_0^\infty\, \alpha_1^A(i \omega)\,
\alpha_2^B(i \omega)\, d\omega \nonumber \\
C^{AB}(2,1) &=& \frac{15}{2\pi}\, \int_0^\infty\, \alpha_2^A(i \omega)\,
\alpha_1^B(i \omega)\, d\omega .
\label{C_12}
\end{eqnarray}

For the Yb--Yb $^1\!S_0 +\, ^3\!P_1^o$ dimer considered in this work,  the expressions for $C_6$ and $C_8$
are more complicated due to the angular dependence,
the $^3\!P^o_1 \rightarrow \, ^1\!S_0$ decay channel and non-vanishing quadrupole moment of the $^3\!P_1^o$ state.
After some transformations, we arrive at the following expression for
the $C_{6}$ coefficient in the $^1\!S_0 +\, ^3\!P_1^o$ case:
\begin{equation}
C_{6}(\Omega _{p})=\sum_{J=0}^{2}A_{J}(\Omega )X_{J},
\label{C6}
\end{equation}
where the angular dependence $A_J(\Omega)$ is represented by
\begin{equation}\label{eq1}
A_{J}(\Omega)=\frac{1}{3}\sum_{\mu =-1}^{1}\left\{ w_{\mu }^{(1)}
\left(
\begin{array}{ccc}
1 & 1 & J \\
-\Omega & -\mu & \Omega +\mu%
\end{array}%
\right) \right\} ^{2}
\end{equation}
with the dipole weights $w_\mu^{(1)}$ given by~\eref{wmu} and $\Omega =0,1$.
It is worth noting that $A_J(\Omega)$ (and, consequently,
the $C_6$ coefficients) do not depend on gerade/ungerade symmetry.

The quantities $X_J$ for the $^1\!S_0 +\, ^3\!P_1^o$ dimer are given by
\begin{equation}
X_J = \frac{27}{2\pi} \int_0^\infty \alpha_1^A(i\omega) \,
\alpha^B_{1J}(i\omega) \, d\omega + \delta X_0 \, \delta_{J,0} .
\label{X_J}
\end{equation}
where $A \equiv \, ^1\!S_0$ and  $B \equiv\, ^3\!P_1^o$ and $\delta X_0$ is defined below.
The possible values of the total angular momentum $J$ are 0, 1, and 2;
$\alpha_1^A(i\omega)$ is the electric-dipole dynamic polarizability of the $^1\!S_0$ state
at the imaginary argument.

The quantity $\alpha^\Phi_{KJ}(i\omega)$ is a part of the
scalar electric-dipole ($K=1$) or electric-quadrupole ($K=2$) dynamic polarizability of
the state $\Phi$, in which the sum over the intermediate states $|n \rangle$ is restricted
to the states with fixed total angular momentum $J_n=J$:
\begin{eqnarray}
&&\alpha^\Phi_{KJ}(i\omega) \equiv \frac{2}{(2K+1)(2J_\Phi+1)}  \nonumber \\
&\times&  \sum_{\gamma_n} \frac{(E_n - E_\Phi)
|\langle \gamma_n, J_n=J ||T^{(K)}|| \gamma_\Phi, J_\Phi \rangle|^2} {(E_n - E_\Phi)^2 + \omega^2} .
\label{alphaB}
\end{eqnarray}
Here, $\gamma_n$ stands for all quantum numbers of the intermediate states
except $J_n$.

The correction $\delta X_0$ to the $X_0$ term in~Eq.(\ref{X_J})
is due to a downward $^3\!P_1^o \rightarrow \,^1\!S_0$ transition
and is given by the following expression:
\begin{eqnarray}
\delta X_0 &=& 2\,|\langle ^{3}\!P_{1}^{o}||d||^{1}\!S_{0}\rangle|^2
\sum_{n\neq \,^3\!P_1^o}
\frac{(E_n-E_{^1\!S_0})\,
|\langle n||d||^1\!S_0 \rangle|^2}{(E_n-E_{^1\!S_0})^2-\omega_0^2} \nonumber \\
&+& \frac{|\langle ^3\!P_1^o||d||^{1}\!S_{0}\rangle |^{4}}{2\omega_0},
\label{delX0}
\end{eqnarray}
where $\omega _{0}\equiv E_{\,^{3}\!P_{1}^{o}}-E_{\,^{1}\!S_{0}}.$

The expression for the $C_8 (^1\!S_0 +\, ^3\!P_1^o)$ coefficient is substantially more complicated, so it
is discussed in the Appendix. 
\section{Method of calculation}
\label{method}
All calculations were carried out by two methods which allows us to estimate the accuracy of the final values.
The first method combines configuration interaction (CI) with many-body perturbation theory
(MBPT)~\cite{DzuFlaKoz96b}. In the second method, which is more accurate, CI is combined with the coupled-cluster
all-order approach (CI+all-order) that treats both core and valence correlation to
all orders~\cite{Koz04,SafKozJoh09,SafKozCla11}.

In both cases, we start from a solution of the Dirac-Fock (DF) equations for the appropriate states of the individual atoms,
\[
\hat H_0\, \psi_c = \varepsilon_c \,\psi_c,
\]
where $H_0$ is the relativistic DF Hamiltonian~\cite%
{DzuFlaKoz96b,SafKozJoh09} and $\psi_c$ and $\varepsilon_c$ are
single-electron wave functions and energies. The calculation was performed
in the V$^{N-2}$ approximation, i.e, the self-consistent procedure was done for
the [$1s^2,...,4f^{14}$] closed core. The B-spline basis set,
consisting of $N=35$ orbitals for each of partial wave with $l\leq5$, was
formed in a spherical cavity with radius 60 a.u.
The CI space is effectively complete. It includes the following orbitals: $6-20s$, $6-20p$,
$5-19d$, $5-18f$, and $5-11g$.

The wave functions and the low-lying energy levels are determined by solving
the multiparticle relativistic equation for two valence electrons~\cite%
{KotTup87},
\begin{equation}
H_{\mathrm{eff}}(E_n) \Phi_n = E_n \Phi_n.
\end{equation}
The effective Hamiltonian is defined as
\begin{equation}
H_{\mathrm{eff}}(E) = H_{\mathrm{FC}} + \Sigma(E),
\end{equation}
where $H_{\mathrm{FC}}$ is the Hamiltonian in the frozen-core approximation.
The energy-dependent operator $\Sigma(E)$ which takes into account virtual
core excitations is constructed using the second-order perturbation theory in
the CI+MBPT method \cite{DzuFlaKoz96b} and using linearized coupled-cluster
single-double method in the CI+all-order approach \cite{SafKozJoh09}.
$\Sigma(E)=0$ in the pure CI approach.
Construction of the effective Hamiltonian in the CI+MBPT and CI+all-order
approximations is described in detail in Refs.~\cite{DzuFlaKoz96b,SafKozJoh09}.
The contribution of the Breit interaction is negligible at the present level of accuracy and
was omitted.

The dynamic polarizability of the $2^K$-pole operator $T^{(K)}$ at imaginary argument is calculated as the sum of
three contributions: valence, ionic core, and $vc$. The $vc$ term subtracts out the ionic core terms which are forbidden by  the Pauli principle.
Then
\begin{equation}
\alpha_K(i\omega) = \alpha_K^v(i\omega) + \alpha_K^c(i\omega) , 
\end{equation}%
where both the core and $vc$ parts are included in $\alpha_K^c(i\omega)$.
\subsection{Valence contribution}
The valence part of the dynamic polarizability, $\alpha_K^{v}(i\omega )$, of
an atomic state $|\Phi \rangle $ is determined by solving the inhomogeneous
equation in the valence space. If we introduce the wave function of
intermediate states $|\delta \Phi \rangle $ as
\begin{eqnarray}
|\delta \Phi \rangle &\equiv &\mathrm{Re}\,\left\{ \frac{1}{H_{\mathrm{eff}%
}-E_{\Phi }+i\omega }\,\sum_{i}|\Phi _{i}\rangle
\langle \Phi _{i}| T^{(K)}_{0} |\Phi \rangle \right\}  \nonumber \\
&=&\mathrm{Re}\,\left\{ \frac{1}{H_{\mathrm{eff}}-E_{\Phi }+i\omega }
\,T^{(K)}_0|\Phi \rangle \right\} ,
\label{delPhi}
\end{eqnarray}%
where ``Re'' means the real part, then $\alpha^{v}(i\omega )$ is given by
\begin{equation}
\alpha^{v}(i\omega )=2\,\langle \Phi |T^{(K)}_0| \delta \Phi \rangle \,.
\label{alphav}
\end{equation}%
Here, $T^{(K)}_0$ is the zeroth component of the $T^{(K)}$ tensor.
We include random-phase
approximation (RPA) corrections to the $2^K$-pole operator $T^{(K)}_{0}$. The
 Eqs.~(\ref{delPhi}) and (\ref{alphav}) can also be used to
find $\alpha_{K J}^{v}$, i.e, the part of the valence polarizability,
where summation goes over only the intermediate states with
fixed total angular momentum $J$. We refer the reader to Ref.~\cite{KozPor99} for further details of this approach.
\subsection{Core contribution}
The core and $vc$ contributions to multipole polarizabilities are evaluated in the single-electron relativistic RPA
approximation. The small $\alpha^{vc}$ term is calculated by adding $vc$
contributions from the individual electrons, i.e., $\alpha^{vc}(6s^2)=2\,\alpha^{vc}(6s)$
and $\alpha^{vc}(6s6p)=\alpha^{vc}(6s)+\alpha^{vc}(6p)$.

A special consideration is required when we need to find the core contribution
to $\alpha^\Phi_{K J}(i\omega)$ of a state $\Phi$.
If we disregard possible excitations of the core electrons to the occupied valence
shells, the valence and core subsystems can be considered as independent.
Then, the total angular momenta ${\bf J}_\Phi$ and ${\bf J}_n$ of the states $\Phi$ and
$\Phi_n$, respectively, can be represented as the sum of the valence and core parts ${\bf J} = {\bf J}^v + {\bf J}^c$.
\begin{table*} [ht]
\caption{Theoretical and experimental~\cite{RalKraRea11} energy
levels (in cm$^{-1}$). Two-electron binding energies are given in the
first row, energies in other rows are counted from the ground state.
Results of the CI, CI+MBPT, and CI+all-order calculations are given
in columns labeled ``CI'', ``CI+MBPT'', and ``CI+All''. Corresponding
relative differences of these three calculations with the experiment
are given in cm$^{-1}$ and in percentages.}
\label{tabE}
\begin{ruledtabular}
\begin{tabular}{rrccrrccrcc}
\multicolumn{1}{c}{\multirow{2}{*}{State}} &
\multicolumn{1}{c}{\multirow{2}{*}{Exper.}} &
\multicolumn{1}{c}{\multirow{2}{*}{CI}} &
\multicolumn{1}{l}{\multirow{2}{*}{CI+MBPT}} &
\multicolumn{1}{c}{\multirow{2}{*}{CI+All}} &
\multicolumn{3}{c}{Differences (cm$^{-1}$)}
& \multicolumn{3}{c}{Differences (\%)} \\
&&&&&\multicolumn{1}{c}{CI} & \multicolumn{1}{c}{CI+MBPT} &
\multicolumn{1}{c}{CI+all}& \multicolumn{1}{r}{CI}
& \multicolumn{1}{r}{CI+MBPT} & \multicolumn{1}{c}{CI+All} \\
\hline
$6s^2\,\,^1\!S_0 $   & 148650& 137648& 150532& 149751&$-$11003 & 1882  & 1101 &$-$7.4 &   1.3 &   0.7\\
$5d6s\,\,^3\!D_1 $   & 24489 & 25505 & 25301 & 25108 &    1016 &  812  & 619  &   4.1 &   3.3 &   2.5\\
$5d6s\,\,^3\!D_2 $   & 24752 & 25522 & 25587 & 25368 &     770 &  835  & 616  &   3.1 &   3.4 &   2.5\\
$5d6s\,\,^3\!D_3 $   & 25271 & 25597 & 26172 & 25891 &     326 &  901  & 620  &   1.3 &   3.6 &   2.5\\
$5d6s\,\,^1\!D_2 $   & 27678 & 25944 & 28842 & 28353 & $-$1734 & 1164  & 676  &$-$6.3 &   4.2 &   2.4\\
$6s7s\,\,^3\!S_1 $   & 32695 & 29631 & 33170 & 33092 & $-$3064 &  475  & 397  &$-$9.4 &   1.5 &   1.2\\
$6s7s\,\,^1\!S_0 $   & 34351 & 31346 & 34848 & 34755 & $-$3005 &  497  & 404  &$-$8.7 &   1.4 &   1.2\\[0.2pc]
$6s6p\,\,^3\!P^o_0 $ & 17288 & 14032 & 18258 & 17760 & $-$3256 &  969  & 472  &$-$19  &   5.6 &   2.7\\
$6s6p\,\,^3\!P^o_1 $ & 17992 & 14675 & 18949 & 18450 & $-$3317 &  957  & 458  &$-$18  &   5.3 &   2.5\\
$6s6p\,\,^3\!P^o_2 $ & 19710 & 16137 & 20698 & 20251 & $-$3574 &  987  & 541  &$-$18  &   5.0 &   2.7\\
$6s6p\,\,^1\!P^o_1 $ & 25068 & 23888 & 26461 & 25967 & $-$1181 & 1393  & 899  &$-$4.7 &   5.6 &   3.6\\
$6s7p\,\,^3\!P^o_0 $ & 38091 & 34649 & 38672 & 38504 & $-$3441 &  581  & 413  &$-$9.0 &   1.5 &   1.1\\
$6s7p\,\,^3\!P^o_1 $ & 38174 & 34736 & 38745 & 38572 & $-$3438 &  571  & 398  &$-$9.0 &   1.5 &   1.0\\
$6s7p\,\,^3\!P^o_2 $ & 38552 & 35045 & 39127 & 38962 & $-$3507 &  575  & 410  &$-$9.1 &   1.5 &   1.1\\
$6s7p\,\,^1\!P^o_1 $ & 40564 & 35697 & 39534 & 39311 & $-$4867 &$-$1030&$-$253&$-$12  &$-$2.5 &$-$3.1\\
\end{tabular}
\end{ruledtabular}
\end{table*}
In our consideration, the core of the $\Phi$ state consists of the closed shells,
and $J_{\Phi}^c=0$. If we assume that the electrons are excited from the core,
while the valence part of the wave function remains the same,
we can express the reduced matrix element of the operator $T^{(K)}$ as
\begin{eqnarray}
&&\langle J_\Phi ||T^{(K)}|| J_n \rangle = \nonumber \\
&&\langle J_{\Phi}^c=0, J_{\Phi}^v, J_{\Phi} ||T^{(K)}|| J_n^c=K, J_{\Phi}^v, J_n \rangle .
\end{eqnarray}
If $T^{(K)}$ acts only on the core part of the system, we arrive at (see, e.g.,~\cite{VarMosKhe88})
\begin{eqnarray}
&&\langle J_{\Phi}^c = 0, J_{\Phi}^v, J_\Phi ||T^{(K)}|| J_n^c=K, J_{\Phi}^v, J_n \rangle \nonumber \\
& = & \sqrt{\frac{2J_n+1}{2K+1}} \langle J_{\Phi}^c=0||T^{(K)}|| J_n^c=K\rangle .
\end{eqnarray}
Then, using~\eref{alphaB}, we can write the core contribution to $\alpha_{KJ}(i\omega)$
of the $\Phi$ state as
\begin{eqnarray}
&&\alpha^c_{KJ}(i\omega) = \frac{2\, (2J+1)}{(2K+1)^2\,(2J_\Phi+1)}  \nonumber \\
&\times&  \sum_{\gamma_n^c} \frac{(E_n - E_\Phi)
|\langle J_{\Phi}^c=0||T^{(K)}|| J_n^c=K\rangle|^2} {(E_n - E_\Phi)^2 + \omega^2} .
\label{alphaPhi}
\end{eqnarray}
Taking into account that the core
polarizability $\alpha^c_K(i\omega )$ of the operator $T^{(K)}$ in a single-electron approximation can be written as
\begin{eqnarray}
\alpha^c_K(i\omega) &=& \frac{2}{2K+1} \nonumber \\
&\times& \sum_{a,n}\frac{\varepsilon_n-\varepsilon_a}{(\varepsilon_n-\varepsilon_a)^{2} + \omega^2}
|\langle n||T^{(K)}||a \rangle|^2 ,
\end{eqnarray}
where $|a\rangle$ and $|n\rangle$ are the single-electron core and virtual states,
we arrive at
\begin{eqnarray}
\alpha^c_{K J}(i\omega )=\frac{2J+1}{(2K+1)(2J_\Phi+1)} \alpha^c_K(i\omega ) .
\end{eqnarray}
Finally, $\alpha_{K J}(i\omega )$ of the $\Phi$ state can be approximated as
\begin{eqnarray}
\alpha_{K J}(i\omega ) = \alpha^{v}_{K J}(i\omega )
+ \frac{2J+1}{(2K+1)(2J_\Phi+1)} \alpha^c_K(i\omega ) ,
\label{alph_KJ}
\end{eqnarray}
where possible values of $J$ are from min$(0,|J_\Phi-K|)$ to $J_\Phi+K$.
\section{Results and discussion}
\label{results}
\subsection{Energy levels}
\label{E_levels}
\begin{figure}
  \includegraphics[width=3.5in]{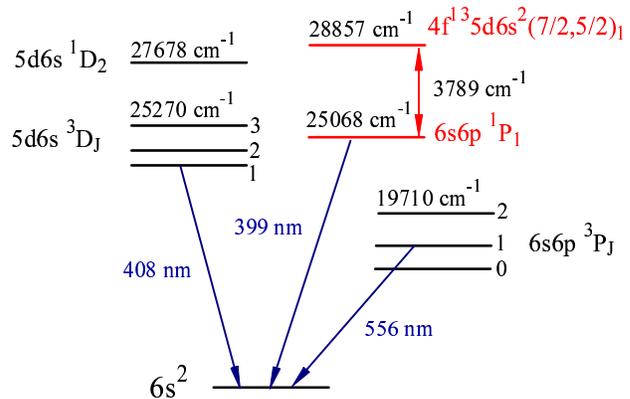}
  \caption{(Color online) Low-lying energy levels of Yb.
   Other states of the  $4f^{13}5d6s^2$ configuration are not shown.}
  \label{fig1}
\end{figure}

We start from the calculation of the low-lying energy levels of atomic Yb.
The calculations were carried out using CI, CI+MBPT, and CI+all-order methods.
The results are listed in~\tref{tabE} (see also the Supplemental Material to Ref.~\cite{SafPorCla12})
in columns labeled ``CI'', ``CI+MBPT'', and ``CI+All''.
Two-electron binding energies are given in the first row, energies in other rows are counted from
the ground state. Corresponding relative differences of these three calculations with experiment
are given in cm$^{-1}$ and in percentages.
The even- and odd-parity levels are schematically presented in Fig.~\ref{fig1}.
\begin{table}
\caption{A breakdown of the contributions to the $6s^2\,^1\!S_0$, $6s6p\,^3\!P^o_0$, and $6s6p\,^3\!P^o_1$ electric-dipole,
$\alpha_1$, and electric-quadrupole, $\alpha_2$, static polarizabilities in the CI+all-order approximation.
For the $^3\!P^o_1$ state, the scalar polarizabilities are given.
The row labeled ``Other'' gives the contribution of all other valence states not explicitly listed in the table.
The row  labeled ``Core+vc'' gives the contributions from the core and $vc$ terms. The row labeled ``Total'' lists the
final values obtained as the sum of all contributions. $|\langle n ||T^{(K)}|| m \rangle|$ are the reduced
matrix elements; $T^{(1)}=d$ and $T^{(2)}=Q$ stand for the electric-dipole and electric-quadrupole
operators, respectively. The theoretical and experimental transition energies are presented in columns
$\Delta E_{\rm th}$ and $\Delta E_{\rm exp}$ (in cm$^{-1}$). The contributions to the polarizabilities
are given in the column labeled ``$\alpha$''.}
\label{brk}
\begin{ruledtabular}
\begin{tabular}{llcrrr}
\multicolumn{1}{c}{Polarizability}
                       &\multicolumn{1}{c}{Contrib.}     &\multicolumn{1}{c}{$|\langle n ||T^{(K)}|| m \rangle|$}
                                                                     &\multicolumn{1}{l}{$\Delta E_{\rm th}$}
                                                                             &\multicolumn{1}{l}{$\Delta E_{\rm exp}$}
                                                                                     &\multicolumn{1}{r}{$\alpha$} \\
\hline
$\alpha_1(^3\!P_0^o)$  &   $5d6s\, ^3\!D_1$  &  2.89   &  7346 &  7201 &   166 \\
                       &   $6s7s\, ^3\!S_1$  &  1.95   & 15332 & 15406 &    36 \\
                       &   $6s6d\, ^3\!D_1$  &  1.84   & 22490 & 22520 &    22 \\
                       &   Other                           &         &       &       &    63 \\
                       &   Core + vc                       &         &       &       &     6 \\
                       &   Total                           &         &       &       &   293 \\[0.4pc] $\alpha_{1s}(^3\!P_1^o)$
                       &   $6s^2\, ^1\!S_0$  &  0.571  &-18450 &-17992 &    -1 \\
                       &   $5d6s\, ^3\!D_1$  &  2.51   &  6656 &  6497 &    46 \\
                       &   $5d6s\, ^3\!D_2$  &  4.35   &  6916 &  6760 &   133 \\
                       &   $5d6s\, ^1\!D_2$  &  0.453  &  9899 &  9686 &     1 \\
                       &   $6s7s\, ^3\!S_1$  &  3.46   & 14642 & 14703 &    40 \\
                       &   $6s7s\, ^1\!S_0$  &  0.243  & 16305 & 16359 &     0.2 \\
                       &   $6s6d\, ^3\!D_1$  &  1.62   & 21800 & 21817 &     6 \\
                       &   $6s6d\, ^3\!D_2$  &  2.78   & 21831 & 21846 &    17 \\
                       &   $6s6d\, ^1\!D_2$  &  0.614  & 22066 & 22070 &     1 \\
                       &   Other                           &         &       &       &    66 \\
                       &   Core + vc                       &         &       &       &     6 \\
                       &   Total                           &         &       &       &   315 \\[0.4pc]
$\alpha_2(^1\!S_0)$    &   $5d6s\, ^3\!D_2$    &  3.00   & 25366 & 24752 &    31 \\
                       &   $5d6s\, ^1\!D_2$    &  25.00  & 28349 & 27678 &  1936 \\
                       &   $6s6d\, ^3\!D_2$    &  3.77   & 40281 & 39838 &    31 \\
                       &   $6s6d\, ^1\!D_2$    &  8.06   & 40516 & 40062 &   141 \\
                       &   Other                           &         &       &       &   407 \\
                       &   Core + vc                       &         &       &       &    14 \\
                       &   Total                           &         &       &       &  2559 \\[0.4pc]

$\alpha_2(^3\!P_0^o)$  &   $6s6p\, ^3\!P^o_2$& 21.60   &  2490 &  2422 & 16449 \\
                       &   $6s7p\, ^3\!P^o_2$& 10.14   & 20202 & 21263 &   447 \\
                       &   Other                           &         &       &       &  3691 \\
                       &   Core + vc                       &         &       &       &    14 \\
                       &   Total                           &         &       &       & 20602 \\[0.4pc]

$\alpha_{2s}(^3\!P_1^o)$
                       &   $6s6p\, ^3\!P^o_2$& 32.69   &  1800 &  1718 & 17372 \\
                       &   $6s6p\, ^1\!P^o_1$&  5.62   &  7517 &  7076 &   123 \\
                       &   $6s7p\, ^3\!P^o_1$&  9.35   & 20122 & 20099 &   127 \\
                       &   $6s7p\, ^3\!P^o_2$& 16.30   & 20512 & 20560 &   379 \\
                       &   $6s7p\, ^1\!P^o_1$&  4.58   & 20861 & 22572 &    29 \\
                       &   Other                           &         &       &       &  3973 \\
                       &   Core + vc                       &         &       &       &    14 \\
                       &   Total                           &         &       &       & 22017 \\
\end{tabular}
\end{ruledtabular}
\end{table}

Table~\ref{tabE} illustrates that the difference between
the theory and the experiment are as large as 19\% for the odd-parity states at the CI stage. When we include the core-core
and core-valence correlations in the second order of the perturbation theory (CI+MBPT method),
the accuracy significantly improves. Further improvement is achieved when we use the CI+all-order method including
correlations in all orders of the MBPT.
 \subsection{Polarizabilities}
In~\tref{brk} we give a breakdown of the main contributions from the intermediate states to the static electric-dipole and electric-quadrupole
polarizabilities of the $6s^2\,^1\!S_0$, $6s6p\,^3\!P^o_0$, and $6s6p\,^3\!P^o_1$ states in the CI+all-order approximation.
For the $^3\!P^o_1$ state the contributions to the scalar parts of the polarizabilities are presented. While we do not explicitly use the sum-over-states to calculate the polarizabilities, we can separately compute contributions of individual intermediate states.
The row labeled ``Other'' lumps contributions of all other valence states not explicitly listed in the table. The row
labeled ``Core+vc'' gives the contributions from the core and $vc$ terms and the row labeled ``Total'' is the
final value obtained as the sum of all contributions.
The theoretical and experimental transition energies are presented in columns
$\Delta E_{\rm th}$ and $\Delta E_{\rm exp}$ (in cm$^{-1}$). We used the theoretical energies when calculating the contributions
of the individual terms to the polarizabilities. These contributions as well the total values of the polarizabilities
are given in the column labeled ``$\alpha$''.

The role of different contributions to the $6s6p\,^3\!P^o_0$ polarizability was analyzed in Ref.~\cite{SafPorCla12} (see the Supplemental Material).
We compare the $^3\!P^o_0$ case with the contributions to the scalar part of
the $^3\!P^o_1$ polarizability given in~\tref{brk}.
We find that the main contributions to the $6s6p\,^3\!P^o_0$ and $6s6p\,^3\!P^o_1$ polarizabilities
are similar in every respect. In particular,
the $5d6s\, ^3\!D_J$ states contribute $\sim$ 57\% to both polarizabilities. The contributions of the
$6s6d\, ^3\!D_J$ states are at the level of 7-10\%. The higher-excited states not explicitly listed in the table,
labeled as ``Other'', contribute $\sim$ 21\% in both cases.

To the best of our knowledge, there are no experimental data for the electric-quadrupole
polarizabilities listed in the table or any transitions that give dominant contributions to $\alpha_2$.
For instance, the main contribution (76\%) to $\alpha_2(^1\!S_0)$ comes from the $5d6s\, ^1\!D_2$ state.
Any accurate experimental data for the $5d6s\,\, ^1\!D_2$ state (lifetime, oscillator strengths, etc)
would  provide an important  benchmark relevant to the ground state quadrupole polarizability.

We also give the breakdown of the $6s6p\, ^3\!P^o_0$ and the scalar part
of $6s6p\, ^3\!P^o_1$ electric-quadrupole polarizabilities. The main contribution (80\%)
comes from the $6s6p\, ^3\!P^o_2$ state in both cases. We note that the remainder contribution
(listed in rows ``Other'') is significant for all polarizabilities considered here.
These contributions are at the level of 15--18\%.
The uncertainties of the polarizability values are discussed later in Section~\ref{unc}.

\begin{table}[tbp]
\caption{The values of the
$D \equiv |\langle 6s6p\,^{3}\!P_{1}^{o}||d||6s^{2}\,^{1}\!S_{0}\rangle |$
matrix element (in a.u.) and $C_{3}$ coefficients in the CI+MBPT and CI+all-order approximations.}
\label{C_3}%
\begin{ruledtabular}
\begin{tabular}{cddd}
           & \multicolumn{1}{c}{CI+MBPT}
                      & \multicolumn{1}{c}{CI+all-order}
                                & \multicolumn{1}{r}{Experiment}  \\
\hline
    $D$    &  0.581   &  0.572  & 0.549(4)\footnotemark[1] \\
           &          &         & 0.5407(15)\footnotemark[2] \\[0.2pc]
$C_3(0_u)$ &  0.225   &  0.218  & 0.1949(11)\footnotemark[2] \\
$C_3(0_g)$ & -0.225   & -0.218  & \\
$C_3(1_u)$ & -0.113   & -0.109  & \\
$C_3(1_g)$ &  0.113   &  0.109  & 0.09685\footnotemark[3]  \\
\end{tabular}
\end{ruledtabular}\begin{flushleft}
\footnotemark[1]{Reference~\cite{BowBudFre99}. The experimental number
was obtained from the weighted $^3\!P_1^o$ lifetime $\tau(^3\!P_1^o)=845 (12)$ ns};\\
\footnotemark[2]{Reference~\cite{BorCiuJul09l} (this error is pure statistical)};
\footnotemark[3]{Reference~\cite{TakSaiTak12l}}.\end{flushleft}
\end{table}

\subsection{$C_3$ coefficients}
The values of the $C_{3}$ coefficients obtained in the CI+MBPT and
CI+all-order approximations for the $^1\!S_0 +\, ^3\!P_1^o$ dimer
are given in Table~\ref{C_3} (also see the Supplemental Material~\cite{PorSafDer13}). We calculated the
$|\langle 6s6p\,^3\!P_1^o||d||6s^2\,^1\!S_0\rangle |$ matrix element (ME)
and then found $C_{3}$ coefficients using Eq.~(\ref{C3detail}).
The $C_3(0_g)$ and $C_3(1_u)$ have the same  numerical values
 as $C_3(0_u)$ and $C_3(1_g)$, but the opposite sign.
Our CI+all-order value for this ME differs from the experimental results
by 4-5\%. It is not unexpected, because the $^1\!S_0 -\, ^3\!P_1^o$ transition is an
intercombination transition and due to cancelation of different contributions
its amplitude is relatively small. It may be also affected by the mixing with the core-excited states that
are outside of our CI space as is discussed in detail in \cite{SafPorCla12}.
As a result, the accuracy of calculation of such MEs
is lower. Using Eq.~(\ref{C3detail}) we can estimate the accuracy of $C_{3}$ coefficients
at the level of 8-10\%.
\subsection{$C_6$  and $C_8$ coefficients}
To find the van der Waals coefficients for the   $^1\!S_0 +\, ^3\!P_0^o$
and $^3\!P_0^o +\, ^3\!P_0^o$ dimers we computed the dynamic electric-dipole
and electric-quadrupole polarizabilities of the $^1\!S_0$ and $^3\!P_0^o$ states at imaginary frequency and then
used Eqs.~(\ref{vdW}) and (\ref{C_12}). In
practice, we computed the $C^{AB}_6$ coefficients  by approximating the integral (\ref{C6}) by
Gaussian quadrature of the integrand computed on the finite grid of discrete imaginary frequencies
~\cite{BisPip92}. The $C_6$ coefficient for the
$^1\!S_0 +\, ^1\!S_0$ dimer was obtained in Ref.~\cite{SafPorCla12}.
\begin{table}
\caption{A breakdown of the contributions to the $C_6(\Omega)$ coefficient for Yb-Yb $(^1\!S_0 +\, ^3\!P_1^o)$ dimer.
The expressions for $X_J$ and $A_J$ are given by Eqs.~(\ref{eq1},\ref{X_J}). The $\delta X_0$ term is
given separately in the second row; it is included in $J=0$ contribution.  The CI+MBPT and CI+all-order values for $X_J$ are given
in columns labeled ``MBPT'' and ``All''. }
\label{tabA}
\begin{ruledtabular}
\begin{tabular}{lccccccc}
\multicolumn{1}{l}{$J$}&\multicolumn{3}{c}{$X_J$}&\multicolumn{2}{c}{$A_J$}&\multicolumn{2}{c}{$C_6(\Omega)$}\\
\multicolumn{1}{c}{}&\multicolumn{1}{c}{MBPT} &\multicolumn{1}{c}{All}  &\multicolumn{1}{c}{HO}
&\multicolumn{1}{c}{$\Omega=0$}& \multicolumn{1}{c}{$\Omega=1$}&\multicolumn{1}{c}{$\Omega=0$}& \multicolumn{1}{c}{$\Omega=1$}\\
\hline
$0$        &  1107   & 1135   &   2.5\% &4/9  &1/9  & 504  &126 \\
$\delta_0$ &   248   &  253   &   2.0\% &4/9  &1/9  & 112  & 28  \\
$1$        &  4564   & 4480   &  -1.9\% &1/9  &5/18 & 498  &1244 \\
$2$        &  6752   & 6702   &  -0.7\% &11/45&19/90&1638  &1415 \\
Sum          &         &        &         &     &     &  2753 &2814
\end{tabular}
\end{ruledtabular}
\end{table}
\begin{table}
\caption{A breakdown of the contributions to the $C_8(\Omega)$ coefficient for Yb--Yb $(^1\!S_0 +\, ^3\!P_1^o)$ dimer.
The  expressions for $X^{J_\alpha J_\beta}_k$ and $A^{J_\alpha J_\beta}_k$ are given in the Appendix~\ref{app}.
The $\delta X^{11}_1$ term (designated as $\delta^{11}_1$) is given separately in the first row;
it is included in the $X^{11}_1$ contribution. The $\delta X^{20}_2$ term (designated as $\delta^{20}_2$) is given separately in the fifth row;
it is included in the $X^{20}_2$ contribution.  The  CI+all-order values are given for  $X^{J_\alpha J_\beta}_k$
and $C_8$; the relative differences of the CI+all-order and CI+MBPT values are given in columns labeled ``HO'' in \%.
The $+/-$ sign corresponds to the ungerade/gerade symmetry, respectively. }
\label{tabB}
\begin{ruledtabular}
\begin{tabular}{lrccccrr}
\multicolumn{1}{l}{$J_\alpha J_\beta,k$}&\multicolumn{1}{r}{$X^{J_\alpha J_\beta}_k$}&\multicolumn{1}{c}{HO}&
\multicolumn{2}{c}{$A^{J_\alpha J_\beta}_k$}&\multicolumn{2}{c}{$C_8(\Omega)$}\\
\multicolumn{3}{c}{} &\multicolumn{1}{c}{$\Omega=0$}& \multicolumn{1}{c}{$\Omega=1$}&\multicolumn{1}{c}{$\Omega=0$}& \multicolumn{1}{c}{$\Omega=1$}\\

\hline
$\delta^{11}_1$  &  66588 &   0.5\%  &            &                &	       &      \\
11,1             & 107772 &   0.0\%  &    3/5     &  1/5           &64663      & 21554 \\
12,1             & 392687 &  -0.6\%  &    1/15    &  7/15          &26179      & 183254\\
13,1             & 249267 &  -1.1\%  &   43/105   & 31/105         &102081     & 73593\\[0.1pc]
$\delta^{20}_2$  &   6510 &   1.4\%  &            &                &	       &      \\
20,2             &  35061 &   3.5\%  &    3/5     &  1/5           &21037      & 7012\\
21,2             & 142845 &  -0.4\%  &    1/5     &  2/5           &28569      & 57138\\
22,2             & 213240 &   0.7\%  &    9/25    &  8/25          &76766      & 68237\\[0.1pc]
11,3             &   1061 &  -5.6\%  &$\pm\,$3/5  & $\pm\,$1/5     &$\pm\,$637 &$\pm$212\\
22,4             &    550 &  -15\%   &$\pm\,$9/25 & $\pm\,$3/25    &$\pm\,$198 &$\pm\,$66 \\[0.3pc]
 $C_8(\Omega_u)$ &        &          &            &                &320130     & 411067  \\
 $C_8(\Omega_g)$ &        &          &            &                &318461     & 410511\\
\end{tabular}
\end{ruledtabular}
\end{table}

\begin{table*}[tbp]
\caption{The $6s^2\,^1\!S_0$, $6s6p\,^3\!P^o_0$, and $6s6p\,^3\!P^o_1$ electric-dipole,
$\alpha_1$, and electric-quadrupole, $\alpha_2$, static polarizabilities in the
CI+MBPT and CI+all-order approximations (in a.u.). For the $^3\!P^o_1$ state
the scalar parts of the polarizabilities are presented.
The values of $C_6(\Omega_{u/g})$ and $C_8(\Omega_{u/g})$ coefficients for the $A+B$ dimers
in the CI+MBPT and CI+all-order approximations are listed in the second part of the table.
The (rounded) CI+all-order values are taken as final.}
\label{C_68}%
\begin{ruledtabular}
\begin{tabular}{llccccl}
\multicolumn{1}{c}{Level}& \multicolumn{1}{c}{Property}&
\multicolumn{1}{c}{CI+MBPT}& \multicolumn{1}{c}{CI+all}&\multicolumn{1}{c}{HO}&
\multicolumn{1}{c}{Final}& \multicolumn{1}{l}{Other}\\
\hline
$6s^2\,^1\!S_0$   &$\alpha_1^{\rm ~a}$     &  138.3  &  140.9  & 1.8\%&  141(2)    & 141(6)$^{\rm b}$ \\
                  &                        &         &         &      &            & 136.4(4.0)$^{\rm c}$  \\
                  &                        &         &         &      &            & 144.59$^{\rm d}$ \\
$6s6p\,^3\!P_0^o$  &$\alpha_1^{\rm ~a}$    &  305.9  &  293.2  &-4.3\%&   293(10)  & 302(14)$^{\rm b}$ \\
$6s6p\,^3\!P_1^o$ &$\alpha_{1s}$           &  323.3  &  315.3  &-2.5\%&   315(11)  &\\
$6s^2\,^1\!S_0$   &$\alpha_2$              &  2484   &  2559   & 2.9\%& 2560(80)   & \\
$6s6p\,^3\!P_0^o$ &$\alpha_2$              & 21294   & 20601   &-3.4\%& 20600(700) & \\
$6s6p\,^3\!P_1^o$ &$\alpha_{2s}$           & 22923   & 22017   &-4.1\%& 22000(900) &\\ [0.4pc]
$^1\!S_0 +\, ^1\!S_0$     & $C_6^{\rm ~a}$    &   1901&   1929   &1.5\%   & 1929(39)          & 1932(35)$^{\rm e}$ \\
                          & $C_8$            & 182360& 187860   &2.9\%   & 1.88(6)$\times 10^5$& 1.9(5)$\times 10^5$$^{\rm e}$\\[0.3pc]
$^1\!S_0 +\, ^3\!P_0^o$   & $C_6$            &   2609&   2561   &-1.9\%  &  2561(95)          & 2709(338)$^{\rm b}$ \\[0.3pc]
$^3\!P_0^o +\, ^3\!P_0^o$ & $C_6$            &   3916&   3746   &-4.5\%  &  3746(180)         & 3886(360)$^{\rm b}$ \\[0.3pc]
$^1\!S_0 +\, ^3\!P_1^o$   & $C_6(0_{u\!/g})$ &   2649&   2640   &-0.3\%  & 2640(103)          & 2410(220)$^{\rm f}$ \\
                          & $C_6(1_{u\!/g})$ &   2824&   2785   &-1.4\%  & 2785(109)          & 2283.6$^{\rm g}$ \\[0.2pc]
                          & $C_8(0_u)$       & 321097& 320130   &-0.3\%  & 3.20(14)$\times 10^5$  &                   \\
                          & $C_8(1_u)$       & 412779& 411067   &-0.4\%  & 4.11(18)$\times 10^5$  &                   \\
                          & $C_8(0_g)$       & 319300& 318461   &-0.3\%  & 3.18(14)$\times 10^5$  &                   \\
                          & $C_8(1_g)$       & 412180& 410511   &-0.4\%  & 4.11(18)$\times 10^5$  &
\end{tabular}
\end{ruledtabular}
\begin{flushleft}
$^{\rm a}${\citet{SafPorCla12}, theory.}\\
$^{\rm b}${\citet{DzuDer10}, theory.}\\
$^{\rm c}${\citet{ZhaDal07}, based on experiment.}\\
$^{\rm d}${\citet{SahDas08}, theory.}\\
$^{\rm e}${\citet{KitEnoKas08l}, experiment.}\\
$^{\rm f}${\citet{BorCiuJul09l}, experiment; the error includes only uncertainty of the fit.}\\
$^{\rm g}${\citet{TakSaiTak12l}, experiment.}
%
%
\end{flushleft}
\end{table*}

The calculation of the $C_6$ and $C_8$ coefficients for the $^1\!S_0 +\, ^3\!P_1^o$ dimer
was carried out according to the expressions given by Eqs.~(\ref{C6})-(\ref{X_J}) and in the Appendix~\ref{app}.
A breakdown of the contributions to the $C_6(\Omega)$ coefficient for Yb--Yb $(^1\!S_0 +\, ^3\!P_1^o)$ dimer is given in Table~\ref{tabA}.
We list the quantities $X_J$ and coefficients $A_J$  given by Eqs.~(\ref{eq1}) and (\ref{X_J})
for allowed $J=0,1,2$. The $\delta X_0$ term is presented separately in the second row to illustrate the magnitude of this contribution.
It is relatively small, 4\% of the total for $\Omega=0$ and 1\% for $\Omega=1$.  It is included in the $X_0$ value given in the table.
We note that the $C_6 (^1\!S_0 +\, ^3\!P_1^o)$ coefficient do not depend on $u/g$ symmetry.
The CI+MBPT and CI+all-order values for $X_J$ are given in columns labeled ``MBPT'' and ``All''.
The relative differences between these values, which give an estimate of the
higher-order contributions, are listed in the column labeled ``HO''. We find that the higher orders contribute with a different sign to $J=0$ and $J=1, 2$.

A breakdown of the contributions to the $C_8(\Omega)$ coefficients for Yb--Yb $^1\!S_0 +\, ^3\!P_1^o$ dimer is given in Table~\ref{tabB}.   We list
the quantities $X^{J_\alpha J_\beta}_k$ and coefficients $A^{J_\alpha J_\beta}_k$
(the analytical expressions for them are given in the Appendix~\ref{app}).
The $\delta X^{11}_1$ and  $\delta X^{20}_2$ terms are given separately in the first and fifth rows; they are included in the
$X^{11}_1$ and $X^{20}_2$ contributions, respectively.
For calculation of $\delta X^{11}_1$ we used the values $|\langle ^3\!P_1^o||Q||^3\!P_1^o \rangle| = 17.75$ a.u.
and  the static $^1\!S_0$ polarizability $\alpha_1^A(0) = 140.9$ a.u. obtained in the CI+all-order approximation.
The coefficients $A^{11}_3$ and $A^{22}_4$ contain $(-1)^p$, therefore their sign is different for gerade and ungerade symmetry resulting in slightly different values for $C_8(\Omega_u)$ and $C_8(\Omega_g)$. In Table~\ref{tabB}, the $+/-$ sign corresponds to the ungerade/gerade symmetry, respectively.
The  CI+all-order values are given for $X_k^{J_\alpha J_\beta}$ and $C_8$; the relative differences of the CI+all-order and CI+MBPT values are given
in column labeled ``HO'' in \%.

Our final results for polarizabilities and the van der Waals $C_6$ and $C_8$ coefficients are summarized in~Table~\ref{C_68}.

The $6s^2\,^1\!S_0$, $6s6p\,^3\!P^o_0$, and $6s6p\,^3\!P^o_1$ electric-dipole,
$\alpha_1$, and electric-quadrupole, $\alpha_2$, static polarizabilities in the
CI+MBPT and CI+all-order approximations are listed in  a.u.. For the $^3\!P^o_1$ state
the scalar parts of the polarizabilities are presented.
The values of $C_6(\Omega_{u/g})$ and $C_8(\Omega_{u/g})$ coefficients for the $A+B$ dimers
in the CI+MBPT and CI+all-order approximations are listed in the second part of the table.
The (rounded) CI+all-order values are taken as final. The relative contribution of the  higher-order corrections
is estimated as the difference of the CI+all-order and CI+MBPT results, it is listed in column labeled ``HO'' in percent.
\section{Determination of uncertainties}
\label{unc}
We compare frequency-dependent polarizabilities calculated in the CI+MBPT and CI+all-order approximations for all $\omega$
used in our finite grid to estimate the uncertainties of the $C_6$ and $C_8$ coefficients.
We find that the difference between the CI+all-order and CI+MBPT frequency-dependent polarizability values is largest
for $\omega=0$ and decreases significantly with increasing $\omega$.
This is reasonable because for large $\omega$ the main contribution to the polarizability comes from its
core part. But the  core parts are the same for both CI+all-order and CI+MBPT approaches.

Therefore, the fractional uncertainty $\delta C^{AB}(l,L)$ ($l,L=1,2$) may be expressed via
fractional uncertainties in the static multipole polarizabilities of the atoms
$A$ and $B$~\cite{PorDer03},
\begin{eqnarray}
 \delta C^{AB}(l,L)  =
 \sqrt{\left( \delta \alpha_l^A(0) \right)^2 +
       \left( \delta \alpha_L^B(0)\right)^2} .
\label{C_AB_l}
\end{eqnarray}

The absolute uncertainties induced in $C_6^{AB}$ and $C_8^{AB}$ ($A \neq B$) are given by
\begin{eqnarray}
\Delta C^{AB}_6 &=&  \Delta C^{AB}(1,1) , \nonumber \\
\Delta C^{AB}_8 &=& \sqrt{ (\Delta C^{AB}(1,2))^2+  (\Delta C_{AB}(2,1))^2 } .
\label{C_AB}
\end{eqnarray}

The polarizabilities and their absolute uncertainties are presented in~\tref{C_68}.
The uncertainties of the electric-dipole $^1\!S_0$ and $^3\!P^o_0$ polarizabilities were
discussed in detail in Ref.~\cite{SafPorCla12}; the uncertainty of the
$^3\!P^o_0$ polarizability was determined to be 3.4\%.
\tref{tabE} illustrates that the accuracy of calculation of the $^3\!P^o_0$ and $^3\!P^o_1$ energy levels
is practically the same ($\sim 2.5\%$ at the CI+all-order stage).
We use the same method of solving the inhomogeneous equation to determine both the $^3\!P^o_0$ and $^3\!P^o_1$ polarizabilities.
The main contributions to these polarizabilities
are also very similar. Based on these arguments, we assume that the uncertainty of the scalar part of
the $^3\!P^o_1$ polarizability can be estimated at the level of 3.5\%.

Our estimates of the uncertainties of the electric-quadrupole polarizabilities are based on
the differences between the CI+MBPT and CI+all-order values.
Besides that we take into account that in all cases the dominant contribution comes from
the low-lying state which energies we reproduce well (see ~\tref{tabE}).
Based on the size of the higher-order correction,  we assign the uncertainties 3-4\% to these polarizabilities.
These results, as well as the final (recommended) values of the polarizabilities,
are presented in Table~\ref{C_68} (see also Ref.~\cite{PorSafDer13}).

Using Eqs.~(\ref{C_AB_l}) and (\ref{C_AB}) we estimated the fractional uncertainties of the $C_6$ coefficient for the
$^1\!S_0 +\, ^3\!P^o_{0,1}$  dimers at the level of
4--4.5\% . The uncertainty of the $C_8(^1\!S_0 +\, ^1\!S_0)$ coefficient
is 3.2\% and the uncertainties of the $C_8(^1\!S_0 +\, ^3\!P^o_1)$ coefficients
are $\sim$ 4.5\%. The difference of the CI+all-order and CI+MBPT values (4.5\%) is taken as an uncertainty for
the $C_6\, (^3\!P^o_0+\,^3\!P^o_0)$ coefficient.

\section{Conclusion}
\label{concl}
To conclude, we evaluated the electric-dipole and electric-quadrupole static
and dynamic polarizabilities for the $6s^2\,^1\!S_0$, $6s6p\,^3\!P^o_0$, and
$6s6p\,^3\!P^o_1$ states and estimated their uncertainties.
The $C_6$ and $C_8$ coefficients  are evaluated for the Yb-Yb dimers. The uncertainties
of our calculations of the van der Waals coefficients do not exceed 5\%.
Our result $C_8=1.88(6) \times 10^{5}$ for the $^1\!S_0+\,^1\!S_0$ dimer
is in excellent agreement with the experimental value $C_8=1.9(5) \times 10^{5}$~\cite{KitEnoKas08l}.
The quantities calculated in this work allow future benchmark tests of molecular theory and experiment.
Most of these quantities are determined for the first time.  Methodology developed in this
work can be used to evaluate properties of other dimers with excited atoms that have a strong decay channel.
\section*{Acknowledgement}
We thank P. Julienne for helpful discussions.
This research was performed under the sponsorship of the
U.S. Department of Commerce, National Institute of Standards and
Technology, and was supported by the National Science Foundation
under Physics Frontiers Center Grant No. PHY-0822671 and by the
Office of Naval Research. The work of S.G.P. was supported in part by
US NSF Grant No.\ PHY-1212442 and RFBR Grant No.\ 11-02-00943.
 The work of A.D. was supported in part by the US NSF Grant No. PHY-1212482.
\begin{widetext}
\appendix
\section{$C_8$ coefficients for the $^1\!S_0 +\, ^3\!P_1^o$ dimer}
\label{app}
Following formalism of Section~\ref{GenForm}, the $C_8$ coefficient may be expressed as:
\[
\frac{C_8(\Omega_p)}{R^8} = \sum_{A,B \neq \alpha,\beta} \frac{\langle A B |%
\hat{V}_{dq}| \alpha \beta \rangle \langle \alpha \beta |\hat{V}_{dq}| A B
\rangle + (-1)^p \langle A B |\hat{V}_{dq}| \alpha \beta \rangle \langle
\alpha \beta |\hat{V}_{dq}| B A \rangle} {E_\alpha+E_\beta-\mathcal{E}} \, ,
\]
which can be further reduced to:
\[
C_8(\Omega_p)= \sum_{k=1}^4 \sum_{J_\alpha J_\beta} A_k^{J_\alpha
J_\beta}(\Omega_p) X_k^{J_\alpha J_\beta} ,
\]
where
\begin{eqnarray*}
A_1^{J_\alpha J_\beta}(\Omega) &=& \sum_{\mu M_\alpha M_\beta} \left\{
w_{\mu }^{(2)} \left(
\begin{array}{ccc}
J_A & 1 & J_\alpha \\
-M_A & \mu & M_\alpha%
\end{array}
\right) \left(
\begin{array}{ccc}
J_B & 2 & J_\beta \\
-M_B & -\mu & M_\beta%
\end{array}
\right) \right\}^2, \\
X_1^{J_\alpha J_\beta} &=& \sum_{\alpha \beta } \frac{|\langle A||d||\alpha
\rangle |^2 |\langle B ||Q|| \beta \rangle|^2} {E_{\alpha} - E_A + E_{\beta}
- E_B} ;
\end{eqnarray*}

\begin{eqnarray*}
A_2^{J_\alpha J_\beta}(\Omega) &=& \sum_{\mu M_\alpha M_\beta} \left\{
w_{\mu}^{(2)} \left(
\begin{array}{ccc}
J_A & 2 & J_\alpha \\
-M_A & \mu & M_\alpha%
\end{array}
\right) \left(
\begin{array}{ccc}
J_B & 1 & J_\beta \\
-M_B & -\mu & M_\beta%
\end{array}
\right) \right\}^2, \\
X_2^{J_\alpha J_\beta} &=& \sum_{\alpha \beta} \frac{|\langle A ||Q|| \alpha
\rangle|^2 |\langle B ||d|| \beta \rangle|^2} {E_\alpha -E_A + E_\beta - E_B} ;
\end{eqnarray*}

\begin{eqnarray*}
A_3^{J_\alpha J_\beta}(\Omega_p) &=& (-1)^p \sum_{\mu \lambda M_\alpha
M_\beta} (-1)^{J_A-J_\alpha+J_B-J_\beta +1} w_\mu^{(2)} w_\lambda^{(2)} \\
&\times& \left(
\begin{array}{ccc}
J_A & 1 & J_\alpha \\
-M_A & \mu & M_\alpha%
\end{array}
\right) \left(
\begin{array}{ccc}
J_A & 1 & J_\beta \\
-M_A & \lambda & M_\beta%
\end{array}
\right) \left(
\begin{array}{ccc}
J_B & 2 & J_\beta \\
-M_B & -\mu & M_\beta%
\end{array}
\right) \left(
\begin{array}{ccc}
J_B & 2 & J_\alpha \\
-M_B & -\lambda & M_\alpha%
\end{array}
\right) , \\
X_3^{J_\alpha J_\beta} &=& \sum_{\alpha \beta } \frac{\langle A ||d|| \alpha
\rangle \langle \alpha ||Q|| B \rangle \langle B ||Q|| \beta \rangle \langle
\beta ||d|| A \rangle } {E_\alpha - E_A + E_\beta - E_B} ;
\end{eqnarray*}

\begin{eqnarray*}
A_4^{J_\alpha J_\beta}(\Omega_p) &=& (-1)^p \sum_{\mu \lambda M_\alpha
M_\beta} (-1)^{J_A - J_\alpha +J_B -J_\beta +1} w_\mu^{(2)} w_\lambda^{(2)}
\\
&\times& \left(
\begin{array}{ccc}
J_A & 2 & J_\alpha \\
-M_A & \mu & M_\alpha%
\end{array}
\right) \left(
\begin{array}{ccc}
J_A & 2 & J_\beta \\
-M_A & \lambda & M_\beta%
\end{array}
\right) \left(
\begin{array}{ccc}
J_B & 1 & J_\beta \\
-M_B & -\mu & M_\beta%
\end{array}
\right) \left(
\begin{array}{ccc}
J_B & 1 & J_\alpha \\
-M_B & -\lambda & M_\alpha%
\end{array}
\right) , \\
X_4^{J_\alpha J_\beta} &=& \sum_{\alpha \beta } \frac{\langle A ||Q|| \alpha
\rangle \langle \alpha ||d|| B \rangle \langle B ||d|| \beta \rangle \langle
\beta ||Q|| A \rangle} {E_\alpha - E_A + E_\beta - E_B}.
\end{eqnarray*}
\end{widetext}
The total angular momenta $J_\alpha$
and $J_\beta$ of the intermediate states $\alpha$ and $\beta$ are fixed in all of the equations above.

We are interested in the case when $A \equiv\, ^{1}\!S_{0}$ and $B \equiv\,
^3\!P_1^o$. Then, $J_A=0$, $J_B=1$, and $\Omega =M_B = 0,1$.

For $k=1$, we have $\ J_{\alpha }=1$ and $J_{\beta }=1,2,3$. The coefficients $A^{1J_\beta}_1(\Omega)$ are listed in Table~\ref{tabB}. The
quantities $X_1^{1\!J_\beta}$ are given by
\begin{eqnarray}
X_1^{1\!J_\beta} &=& \frac{45}{2\pi}\, \int_0^\infty \alpha^A_1(i\omega)\,
\alpha^B_{2 J_\beta}(i\omega)\, d\omega + \delta X_{1}^{11}\, \delta_{J_\beta, 1}\,  \nonumber \\
\delta X_{1}^{11} &=& \frac{3}{2}\,|\langle ^3\!P_1^o||Q||^3\!P_1^o \rangle|^2 \, \alpha_1^A(0) .
\label{X_1}
\end{eqnarray}

For $k=2$, we have $J_\alpha =2$ and $J_\beta=0,1,2$. The coefficients $A^{2J_\beta}_2(\Omega)$ are listed in Table~\ref{tabB}. The
quantities $X^{2J_\beta}_2$ are given by
\begin{eqnarray}
X_{2}^{2J_\beta} &=& \frac{45}{2\pi }\,\int_{0}^{\infty }
\alpha_2^A(i\omega) \, \alpha^B_{1 J_\beta} (i\omega)\, d\omega +
\delta X_{2}^{20}\, \delta_{J_\beta,0}\,   \\
\delta X_{2}^{20} &=&2\,|\langle ^{3}\!P_{1}^{o}||d||^{1}\!S_{0}\rangle
|^{2}\sum_{n}\frac{(E_{n}-E_{\,^{1}\!S_{0}})\,|\langle
n||Q||^{1}\!S_{0}\rangle |^{2}}{(E_{n}-E_{\,^{1}\!S_{0}})^{2}-\omega _{0}^{2}%
}, \nonumber \label{X_2}
\end{eqnarray}%
where $\omega _{0}\equiv E_{\,^{3}\!P_{1}^{o}}-E_{\,^{1}\!S_{0}}$.

For $k=3$, we find that $J_{\alpha }=1$ and $J_{\beta }=1$. For
all other $J_{\alpha }$ and $J_{\beta }$ this expression turns to zero. Then,%
\begin{eqnarray}
A_{3}^{11}(\Omega_{p}= 0) &=&(-1)^{p}\,3/5, \nonumber \\ 
A_{3}^{11}(\Omega_{p}= 1) &=&(-1)^{p}\,1/5.
\end{eqnarray}

\begin{eqnarray}
X_{3}^{11}&=&\sum_{n,k}\,\langle
^{1}\!S_{0}||d||n\,^{1,3}\!P_{1}^{o}\rangle \langle
n\,^{1,3}\!P_{1}^{o}||Q||^{3}\!P_{1}^{o}\rangle\nonumber \\ &\times&\frac{ \langle
^{3}\!P_{1}^{o}||Q||k\,^{1,3}\!P_{1}^{o}\rangle \langle
k\,^{1,3}\!P_{1}^{o}||d||^{1}\!S_{0}\rangle }{E_{n}-E_{\,^{1}%
\!S_{0}}+E_{k}-E_{\,^{3}\!P_{1}^{o}}}.  \label{X_3}
\end{eqnarray}

For $k=4$, we have $J_{\alpha }=2$ and $J_{\beta }=2$. Then%
\begin{equation}
A_{4}^{22}(\Omega_{p}=0)=(-1)^{p}\,\frac{9}{25}; \hspace{0.2cm}
A_{4}^{22}(\Omega_{p}=1)=(-1)^{p}\,\frac{3}{25} ,\nonumber
\end{equation}

\begin{eqnarray}
X_{4}^{22}&=&\sum_{n,k}\,\langle ^{1}\!S_{0}||Q||n\,^{1,3}D_{2}\rangle
\langle n\,^{1,3}\!D_{2}||d||^{3}\!P_{1}^{o}\rangle \nonumber\\
&\times&\frac{ \langle
^{3}\!P_{1}^{o}||d||k\,^{1,3}\!D_{2}\rangle \langle
k\,^{1,3}\!D_{2}||Q||^{1}\!S_{0}\rangle }{E_{n}-E_{\,^{1}\!S_{0}}+E_{k}-E_{%
\,^{3}\!P_{1}^{o}}}.  \label{X_4}
\end{eqnarray}


\end{document}